\documentclass[aps,prl,twocolumn,superscriptaddress]{revtex4-2}
\usepackage[colorlinks=true,citecolor=blue,urlcolor=blue,linkcolor=blue,pdfstartview=FitH,bookmarksopen]{hyperref}
\usepackage{comment}
\usepackage{graphicx}
\usepackage{tipa}
\usepackage{braket}
\usepackage{amsmath, amssymb, mathtools}
\usepackage{amsfonts}
\usepackage{epsfig,float}

\usepackage{graphicx}   
\usepackage{verbatim}   
\usepackage{color}      

\def\ba{\begin{align}}
	\def\ea{\end{align}}
\def\be{\begin{equation}}
	\def\ee{\end{equation}}

\def\bea{\begin{eqnarray}}
	\def\eea{\end{eqnarray}}

\newcommand{\roughly}[1]{\mathrel{\raise.3ex\hbox{$#1$\kern-0.85em
			\lower1ex\hbox{$\sim$}}}}

\def\d{{\partial}}

\def\cO{{\cal O}}

\def\cL{{\cal L}}

\renewcommand\d{\partial}

\newcommand\p{{\mathbf{p}}}

\renewcommand\Re{\mathop{\mathrm{Re}}}
\renewcommand\Im{\mathop{\mathrm{Im}}}

\renewcommand\d{\partial}
\newcommand\q{\mathbf{q}}
\newcommand\x{\mathbf{x}}
\newcommand\y{\mathbf{y}}
\renewcommand\k{\mathbf{k}}

\renewcommand\Im{\mathop{\mathrm{Im}}}

\newcommand{\beq}{\begin{equation}}
	\newcommand{\eeq}{\end{equation}}
\newcommand\dl{\overset{\leftarrow}\partial}
\newcommand\dr{\overset{\rightarrow}\partial}

\renewcommand\Re{\mathop{\mathrm{Re}}}
\renewcommand\Im{\mathop{\mathrm{Im}}}

\begin{document}
	
	\title{Noncommutative Field Theory of the Tkachenko Mode: Symmetries and Decay Rate}
	
	\author{Yi-Hsien Du}
	\affiliation{Kadanoff Center for Theoretical Physics, University of Chicago, Chicago, IL 60637, USA}
	
	\author{Sergej Moroz}
	\affiliation{Department of Engineering and Physics, Karlstad University, Karlstad,  651 88 Sweden}
		\affiliation{Nordita, KTH Royal Institute of Technology and Stockholm University, Stockholm, 114 19, Sweden}
	\author{Dung Xuan Nguyen}
	\affiliation{Center for Theoretical Physics of Complex Systems, Institute for Basic Science (IBS), Daejeon, Korea, 34126}
	
	\author{Dam Thanh Son}
	\affiliation{Kadanoff Center for Theoretical Physics, University of Chicago, Chicago, IL 60637, USA}

\begin{abstract}
We construct an effective field theory describing the collective Tkachenko oscillation mode of a vortex lattice in a two-dimensional rotating Bose-Einstein condensate in the long-wavelength regime. The theory has the form of a noncommutative field theory of a Nambu-Goldstone boson, which exhibits a noncommutative version of dipole symmetry. From the effective field theory, we show that, at zero temperature, the decay width $\Gamma$ of the Tkachenko mode scales with its energy $E$ as $\Gamma\sim E^3$ in the low-energy limit. We also discuss the width of the Tkachenko mode at a small temperature. 

\end{abstract}
\date{December 2022}

\maketitle

	
\emph{Introduction.}---When a superfluid rotates, a lattice of quantized vortices forms. The oscillations of the vortex lattice, the so-called Tkachenko mode~\cite{tkachenko1965,tkachenko1966,tkachenko1969} (for a recent review, see Ref.~\cite{Sonin:2014}), has many distinctive properties. Unlike ordinary sound waves in a solid, at low momenta, the Tkachenko wave has a quadratic dispersion relation $\omega\sim q^2$ and only one polarization~\cite{Sonin1976, volovik1979, Baym2003}. 
The Tkachenko mode is a consequence of a rather intricate realization of spontaneous symmetry breaking: there are many symmetries broken by the superfluid vortex lattice, but only one Nambu-Goldstone boson (NGB)~\cite{PhysRevLett.110.181601, supp}. The Tkachenko mode should exist in rotating superfluid $^4$He, but it has been observed most conclusively in the rotating Bose-Einstein condensate of ultracold atoms~\cite{Coddington:2003}. At a much larger length scale, the Tkachenko mode has been suggested to be the source of an oscillation mode of the Crab pulsar~\cite{Ruderman:1970}.

As the Tkachenko mode is the only low-energy degree of freedom, one expects that it can be described by an effective field theory (EFT) which involves a single field.  However, up to now, a complete understanding of the structure of such a theory has yet to be achieved.  At the quadratic level, the effective Lagrangian~\cite{PhysRevLett.110.181601} coincides with that for a Lifshitz scalar~\cite{Grinstein:1981rbe}, but the form of the interaction terms in the Lagrangian and how they are constrained by symmetries are not known.  These interaction terms are needed to calculate the decay rate of the Tkachenko mode~\cite{Matveenko:2011}.

In this Letter, we show that noncommutative field theory (see, e.g., Refs.~\cite{Douglas:2001ba,Rubakov-NC}) provides a convenient framework for constructing the effective field theory of the Tkachenko mode.  That noncommutative field theory (NCFT) may be applicable to the problem is intuitively understandable---rotating a nonrelativistic system is formally equivalent to placing it in a magnetic field, and on the lowest Landau level (LLL) the guiding-center coordinates do not commute.  Because of that, NCFT has often been invoked in the context of the quantum Hall effect~\cite{Susskind:2001fb,Polychronakos:2001mi,Hellerman:2001rj,Fradkin:2002qw,Pasquier:2007nda,Dong:2020bkt,Du:2021hes}.
Vortex lattices can also be realized on the LLL~\cite{Ho:2001, Sinova:2002, PhysRevA.72.021606}.
As we will see, in the case of the Tkachenko mode, NCFT provides a way to organize terms in the Lagrangian consistent with symmetries.  Following the formalism, we are able to determine the general structure of the interacting Lagrangian, and from there, that the decay rate of a Tkachenko mode (at zero temperature) scales like the cube of its energy,
\begin{equation}
  \Gamma \sim E^3.
\end{equation}
This implies, in particular, that the Tkachenko mode becomes a more and more well-defined quasiparticle (i.e., $\Gamma/E\to 0$) as the energy $E$ approaches zero.

We will also establish a connection between the Tkachenko mode and the ``dipole'' symmetry, which recently became a popular topic (see, e.g., Refs~\cite{Pretko:2016kxt,PhysRevLett.120.195301,Nandkishore:Review,gromov2018towards,Wang:HR,Pretko:FractonReview,Gromov:2020yoc,Seiberg:2020bhn,Nguyen:VLF,Lake:2022ico,Gorantla:2022eem,Kapustin:2022fzp,Gorantla:2022ssr,radzihovsky2022lifshitz,Gromov:Review,nguyen2023quantum,du2023quantum}).  The Tkachenko mode realizes a more complex version of dipole symmetry: 
the magnetic translations, which form a nonabelian group.

\emph{Tkachenko mode as a noncommutative Nambu-Goldstone boson.}---One can arrive at the theory of the Tkachenko mode from microscopic considerations, taking, for example, as the starting point the microscopic theory of bosons with short-range repulsive interactions and then eliminating all redundant degrees of freedom \cite{PhysRevLett.110.181601}. It is instructive, however, to derive the most general form of the effective Lagrangian, relying solely on symmetries.  Such an approach has the advantage of being applicable for strongly correlated rotating superfluids where microscopic calculations are not reliable, e.g., close to a quantum melting transition of the vortex crystal \cite{Cooper2008}.

We first note that the lattice of vortices can be described, as a two-dimensional solid, by two ($a=1,2$) scalar fields $X^a(t, x^i)$; they present the coordinates frozen in the solid~\cite{Leutwyler1996,soper2008}. In this description, in Cartesian coordinates the lattice displacement $u^a$ is related to $X^a$ by $X^a= x^a - u^a$. The vortex current is related to $X^a$ by 
\begin{equation}
\label{eq:jv}
  j^\mu_v = \frac12 n_0 \epsilon^{\mu\nu\lambda} \epsilon^{ab} \d_\nu X^a \d_\lambda X^b,
\end{equation}
where $n_0$ is the equilibrium vortex density.  For a superfluid under rotation with angular velocity $\Omega$, $n_0=\frac1{2\pi}B$, where the effective magnetic field $B=2m\Omega$ with $m$ being the mass of the elementary boson.

In a superfluid, the vortices carry charge with respect to the $u(1)$ dynamical gauge field $a_\mu$ dual to the superfluid Nambu-Goldstone boson \cite{peskin1978, DasguptaHalperin}.  The boson particle number current is expressed as $j^\mu=\frac1{2\pi}\epsilon^{\mu\nu\lambda}\d_\nu a_\lambda$.  The Lagrangian of the system contains terms that describe the coupling of the vortex current with the dual $u(1)$ gauge field and the kinetic and potential terms of the latter,
\begin{equation}
\label{eq:L0}
\mathcal{L}=  -j_v^\mu a_\mu + \frac{m}{4\pi b} \mathbf{e}^2 
  - \epsilon\left( \frac b{2\pi}\right)
  + \frac1{2\pi}\epsilon^{\mu\nu\lambda}A_\mu\d_\nu a_\lambda,
\end{equation}
where $e_i=\partial_0 a_i-\partial_i a_0$, and $b=\epsilon^{ij}\d_i a_j$. In the above equation, $m\mathbf{e}^2/(4\pi b)$ represents the kinetic energy of the superfluid condensate, $\epsilon(b/2\pi)$ is the internal energy as a function of the density, and $A_\mu$ is the gauge potential of the external effective magnetic field $B$. In the lowest Landau level limit $m\to0$, the kinetic term vanishes.  In fact, this term can be dropped if one is eventually interested in the limit $\omega\sim q^2$: in this regime $\mathbf e^2 \ll b^2$.  Without the $\mathbf{e}^2$ term, variation with respect to $a_0$ give a constraint 
\begin{equation} \label{incompressibility}
  \frac12\epsilon_{ab}\epsilon^{ij}\d_i X^a \d_j X^b=1.
\end{equation}
That means the map from $x^i$ to $X^a$ is area-preserving.  To  linear order in the displacement $u^i$, Eq.~\eqref{incompressibility} implies $\d_i u^i=0$, i.e., the displacement is divergence free: the Tkachenko mode is a transverse sound.

The quadratic theory of this transverse sound is analyzed in the Supplement Material (SM) \cite{supp}.  Here we would like to resolve the constraint \eqref{incompressibility} at the nonlinear level. This can be done iteratively, as worked out in the SM \cite{supp}. Here we take a more elegant approach:
on the LLL, one expects the spatial coordinates $x$ and $y$ to become noncommutative (see e.g., Refs.~\cite{Susskind:2001fb,Pasquier:2007nda}),
\begin{equation}
   [ \hat x, \, \hat y] = i\theta, \quad \theta = -\ell^2.
\end{equation}
where $\ell=1/\sqrt{B}$ is the magnetic length. The quantum version of Eq.~(\ref{incompressibility}) then can be written as
\begin{equation}
  [\hat X, \, \hat Y] = i\theta .
\end{equation}
We then conclude that $\hat X^a$ and $\hat x^a$ are related by a unitary transformation,
\begin{equation}
\label{eq:transX}
  \hat X^a = e^{i\hat\phi} \hat x^a 
  e^{-i\hat\phi} ,
\end{equation}
where the operator $\hat\phi$ is an arbitrary function of the two noncommuting coordinates $\hat x$ and $\hat y$.  In noncommutative field theories~\cite{Douglas:2001ba,Rubakov-NC}, any operator corresponds to a Weyl symbol which is a function in space, and the above equation becomes
\begin{equation}\label{Xx-star}
  X^a(x) = e^{i\phi(x)} \star x^a \star e^{-i\phi(x)} .
\end{equation}
Here the star product is defined as $f\star g \equiv f(x) \exp(\frac i2\theta\epsilon^{ij}\dl_i\dr_j)g(x)$.
To linear order in $\phi$ the displacement $u^a$ is then
\begin{equation}\label{u-phi}
  u^a = \{\!\{ \phi, \, x^a\}\!\} = -\theta \epsilon^{ab} \d_b \phi ,
\end{equation}
where $\{\!\{\cdot,\cdot\}\!\}$ denotes the Moyal bracket, $\{\!\{ f,\, g \}\!\}=2f\sin(\frac12\theta\epsilon^{ij}\dl_i\dr_j)g$.  As expected, to this order, the displacement is purely transverse.  To all orders in $\phi$, Eq.~(\ref{Xx-star}) can be written as
\begin{equation}
  X^a = x^a + i \{\!\{U,\, x^a\}\!\} \star U^{-1}
      = x^a +\theta \epsilon^{ai}D_i \phi,
\end{equation}
where $U=e^{i\phi}$, $U^{-1}=e^{-i\phi}$ and
\begin{equation}\label{Diphi}
  D_i \phi \equiv -i\d_i U \star U^{-1}.
\end{equation}
Thus, we identify the Tkachenko mode with a Nambu-Goldstone boson of a noncommutative field theory.  We now show that this field is a compact scalar that shifts under the particle number U(1) symmetry.

\emph{Magnetic translations as noncommutative dipole symmetry}---On the LLL, translations are magnetic translations and do not commute:
\begin{equation}\label{PPcomm}
  [ \hat P_x, \, \hat P_y] = -\frac i \theta \hat{Q},
\end{equation}
where $\hat{Q}$ denotes the boson particle number operator. In our case, the Tkachenko mode is the only low-energy degree of freedom, so it should provide a nontrivial representation of magnetic translations. In the noncommutative theory, translations are realized as a special class of unitary transformations that are exponents of a linear function of coordinate. Acting on $\hat X^a$, such a transformation changes the Weyl symbol of the latter as
\begin{equation}\label{Xa-alpha}
  X^a \to
  e^{i\alpha_i x^i} \star X^a \star e^{-i\alpha_i x^i}
   = X^a(\vec x-\vec \xi),
\end{equation}
with $\xi^i = -\theta\epsilon^{ij}\alpha_j$.  This is a spatial translation by $\vec \xi$.  Viewing $X^a$ as fields in a field theory, such a translation is supposed to be generated by $\hat X^a\to e^{-i \vec \xi\cdot\hat{\vec P}} \hat X^a e^{i\vec \xi\cdot \hat {\vec P}}$.  
Thus, we can identify the magnetic translation as \footnote{The hat symbol $\hat{\phantom{a}}$ on the two sides of Eq.~(\ref{Pequalx}) have different meanings: on the left-hand side, it denotes an operator of field theory, on the right-hand side, an operator in the sense of a function of noncommutative coordinates.}
\begin{equation}\label{Pequalx}
    \hat P_i = \frac 1{\theta}\epsilon_{ij}\hat x^j.
\end{equation}
According to Eqs.~(\ref{Xx-star}) and (\ref{Xa-alpha}) and the associativity of the star product, magnetic translations by $\vec c$ act on the Tkachenko field $\phi$ as multiplication on the left
\begin{equation}\label{NC-dipole}
  e^{i\phi} \to \exp\left(\frac i{\theta}\epsilon_{ij} c^i x^j\right) \star e^{i\phi} .
\end{equation}
This allows us to interpret the action of magnetic translations on a Tkachenko field as a noncommutative version of a dipole symmetry.
Expanding in $\phi$, Eq.~(\ref{NC-dipole}) reads
\begin{equation}\label{phi-magn-tr}
  \phi \to \phi + \frac1{\theta}\epsilon_{ij} c^i x^j
  -\frac12 c^i \d_i \phi + \cdots .
\end{equation}
To leading order, these are simply a dipole symmetry transformation $\phi\to\phi+\alpha_i x^i$ with the parameter $\alpha_i= \theta^{-1}\epsilon_{ij} c^j$, but in addition, there are an infinite number of terms composed of derivatives acting on fields $\phi$. These terms make the magnetic translations noncommuting, as in Eq.~(\ref{PPcomm}). 

Knowing the transformation law for $\phi$ under magnetic translations, we can find the transformation law for $\phi$ under particle number $U(1)$ symmetry. Apply four translations on $e^{i\phi}$, one after another: $e^{-i\beta \hat P_y}e^{-i\alpha \hat P_x}e^{i\beta \hat P_y}e^{i\alpha \hat P_x}$, from Eq.~(\ref{NC-dipole}) we see that $e^{i\phi}$ becomes $e^{i(\phi+\frac{\alpha\beta}{\ell^2})}$.  But we also know from Eq.~(\ref{PPcomm}) that the product of the four translations is $e^{i\frac{\alpha\beta}{\ell^2} \hat Q}$. Thus, under U(1) charge, the Tkachenko field transforms exactly as the phase of the superfluid condensate: $\phi \to \phi + c$. 
The Tkachenko field, therefore, has a dual role: it is the condensate phase, but at the same time, its gradient is the lattice displacement. 
Such a dual role is possible, of course, because at low energy, the condensate phase is entirely determined by the configuration of the vortex lattice. By the ``condensate phase,'' one should have in mind the regular part of the phase where the singular contributions from the vortices have been subtracted away.

As a condensate phase, $\phi$ then should be a compact scalar field with periodicity $2\pi$: $\phi\sim\phi+2\pi$.  The periodicity of $\phi$ can also be seen from the following argument. Let us put the system on a torus of size $L_x\times L_y$.  Then the magnetic field breaks translation symmetry along the $x$ direction to a discrete group of finite translations generated by $x\to x+2\pi \ell^2/L_y$ (which can be seen by computing the Wilson line of the gauge field along a curve wrapping the torus along the $y$ direction at fixed $x$).  This discrete translation is generated by the operator $e^{2\pi i\hat y/L_y}$ under which $\phi\to\phi+2\pi y/L_y+ \cdots$.  This is allowed only when the identification $\phi\sim \phi+2\pi$ is valid.

\emph{Ingredients for a Lagrangian.}---We now write down the most general Lagrangian consistent with symmetries for the field 
 $U=e^{i\phi}$.  The symmetries include global $U(1)$ phase rotations $U\to e^{i\alpha} U$, global magnetic translations $U\to e^{i\vec\alpha\cdot\vec x} \star U$ (noncommutative dipole symmetry),and global rotation $U\to e^{\frac i2\omega x^2}\star U$.
The structures that are
covariant (i.e., transforming like $O\to e^{i\vec\alpha\cdot\vec x} \star O \star e^{-i\vec\alpha\cdot\vec x}$, etc.)
under these transformations are
\begin{subequations}
\begin{align}
   D_0\phi &\equiv -i \d_t U \star U^{-1}, \\
   D_{ab}\phi & = \frac12 (\d_a D_b\phi + \d_b D_a\phi - \delta_{ab}\d_c D_c\phi)\nonumber \\
   & ~ + \frac\theta4 [\epsilon_{ac}\d_i D_c\phi \star \d_i D_b\phi
  + (a\leftrightarrow b)] , \label{Dij}
\end{align}
\end{subequations}
where $D_i \phi$ is defined as in Eq.~(\ref{Diphi}).
Note that $D_{ab}\phi$ is symmetric and traceless \footnote{One way to obtain the structure~(\ref{Dij}) is to start from $U^{ab}=\d_i X^a \star \d_i X^b$.  To leading order in derivaties, $U^{ab}$ is a symmetric matrix with determinant equal to 1, since the transformation from $x^i$ to $X^a$ is area-preserving, and thus has two independent components. The two components of $D_{ab}\phi$ are then related to $U^{ab}$ by a rotation by $90$ degrees:
$$
  D_{ab} \phi = -\frac1{4\theta} (\epsilon^{ac} U^{cb} + \epsilon^{bc} U^{ac}). 
$$
}.

These can be expanded
infinite series over $\phi$.  These series have the property that, at the order $\phi^n$ with a given integer $n$, the leading terms (in derivatives) have $2n$ derivatives if one count $\d_t$ as two derivatives, $\d_t\sim \d_i^2$. This counting is natural as the Tkachenko mode, which is the only low-energy degree of freedom, has a quadratic dispersion.
Keeping at each power of $\phi$ only terms with the minimal number of derivatives,
we have
\begin{subequations}
\begin{align}
  D_0\phi &= \dot\phi + \frac\theta2 \epsilon^{ij}\d_i\dot\phi \d_j\phi + \cdots,\\
  D_{ab}\phi &= \d_a\d_b\phi + \frac\theta2\epsilon^{kl}\d_a\d_b\d_k\phi\d_l\phi
- \text{trace} \nonumber\\
  & ~~ + \frac\theta4 [\epsilon_{ac}\d_i \d_c\phi\, \d_i \d_b\phi
  + (a\leftrightarrow b)]
  + \cdots.
\end{align}
\end{subequations}


\emph{Effective Lagrangian.}---We can now write down the Lagrangian of the Tkachenko mode, keeping at each power of $\phi$ terms with the minimal number of derivatives, counting each occurrence of $\d_t$ as two derivatives.  This Lagrangian would allow one to compute the rate of any scattering process to leading order over the momenta of the particles involved, similarly to the nonlinear Lagrangians for superfluids~\cite{Greiter:1989qb} or solids~\cite{Leutwyler1996,soper2008}. 
In the SM \cite{supp} we explicitly derive a  nonlinear effective theory of the Tkachenko field $\phi$ from the leading-order effective theory of a vortex lattice introduced in Refs.~\cite{MorozVL:2018, PhysRevLett.122.235301}.

The most general Lagrangian consistent with the $U(1)$ and magnetic translation symmetries is a function of the invariant structures defined above:
\begin{equation}
  L = L \bigl(D_t\phi, D_{ab}\phi\bigr).
\end{equation}
The form of the Lagrangian can be restricted further by imposing additional symmetries. In particular, assuming the vortex lattice is a triangular lattice, one should expect the $C_6$ group of rotations by angles multiple of $\frac{2\pi}6$. Introducing the complex coordinate $z=x+iy$, the rotationally invariant structures are now
\begin{equation}\label{C6-inv}
  D_0\phi,\, (D_{ab}\phi)^2,\, 
  \Re (D_{zz}\phi)^3,\,  \Im (D_{zz}\phi)^3 .
\end{equation}

A system of particles in a magnetic field has an antiunitary $RT$ symmetry that combines spatial reflection ($R$) and time reversal ($T$):
\begin{equation}
  x\to x, \quad y\to -y, \quad t\to -t,\quad i\to -i .
\end{equation}
Under this symmetry, $\phi\to-\phi$, which can be seen from its connection to the displacement $u^a$ in Eq.~(\ref{u-phi}). Among the $C_6$ invariants in Eq.~(\ref{C6-inv}), $\Re (D_{zz}\phi)^3$ is odd, while the rest are even.
Thus the most general effective Lagrangian is a function of four arguments,
\begin{equation}
  L = L \Bigl( D_0\phi,\, (D_{ab}\phi)^2, \,
  \Im (D_{zz}\phi)^3, \,
  \bigl(\Re (D_{zz}\phi)^3\bigr)^2 \Bigr).
\end{equation}

\emph{The Girvin-MacDonald-Platzman (GMP) algebra.}---The NCFT construction realizes a key feature of the LLL---the GMP algebra~\cite{GMP}.  Indeed, upon canonical quantization, the particle number density $n=-\delta S/\delta(D_0\phi)$ realizes the NC U(1) gauge transformation, i.e.,
\begin{equation}
  \left[\int\!d^2y\, \lambda(y) n(y), \, O(x) \right] =
  i\delta_\lambda O(x),
\end{equation}
where $\delta_\lambda O$ is the infinitesimal change of $O$ under the gauge transformation, under which $e^{i\phi}\to e^{i\lambda}\star e^{i\phi}$.  But the gauge transformations do not commute: $[\delta_\alpha, \, \delta_\beta]= \delta_{\{\!\{\alpha,\, \beta\}\!\}}$.  From this, one derives the GMP algebra satisfied by $n(x)$.  This is confirmed by explicit calculation in the SM \cite{supp}.

\emph{Quadratic Lagrangian.}--The only terms that contribute to the quadratic Lagrangian are $(D_0\phi)^2$ and $(D_{ij}\phi)^2$.  Modulo a total derivative, the quadratic Lagrangian is that of the quantum Lifshitz model~\cite{Grinstein:1981rbe}
\begin{equation}\label{L2}
  \mathcal L_2 = \frac{c_0}2 (\d_0\phi)^2 - \frac{c_1}2 (\nabla^2\phi)^2,
\end{equation}
which corresponds to a quadratic dispersion relation $\omega\sim q^2$, see also the SM \cite{supp} for the explicit expression for the coefficients $c_0$ and $c_1$.  
This quadratic dispersion relation is protected by the magnetic translation symmetry \cite{PhysRevLett.110.181601}.  From Eq.~(\ref{L2}), one easily reproduces the power-law behavior of the correlation function of the superfluid order parameter at large distances, first found in Ref.~\cite{Sinova:2002} (see also Refs.~\cite{Pitaevskii:1993,Lake:2022ico,Kapustin:2022fzp}).

\emph{Decay width of the Tkachenko mode.}---The quadratic dispersion relation of the Tkachenko mode allows a decay of one Tkachenko quantum into two quanta.  To find the rate of such decay, we need to determine the interaction vertices cubic in the field $\phi$.  It is easy to see that, even as cubic terms appear when one expands the ``quadratic'' terms $(D_0\phi)^2$ and $(D_{ij}\phi)^2$ to cubic order in $\phi$, these terms are total derivatives.  The real cubic interaction appears from the following terms in the Lagrangian: $(D_0\phi)^3$,
$D_0\phi, (D_{ij}\phi)^2$, and $\Im (D_{zz}\phi)^3$.
Up to a total derivative, the cubic Lagrangian has the form
\begin{equation}
  \mathcal L_3 = g_1 (\d_0\phi)^3 + g_2 (\d_0\phi)(\nabla^2\phi)^2 + g_3 \Im (\d_z\d_z\phi)^3.
\end{equation}
From this, one easily finds the energy dependence of the decay width of the Tkachenko mode.  All the cubic interaction terms scale the same way in the scaling scheme with $\d_0 \sim \d_i^2$.  In this scheme, $\phi$ is dimensionless and the $g$'s have dimension $p^{-2}\sim E^{-1}$. 
The decay width $\Gamma$ is proportional to $g^2$, and to have the correct dimension, $\Gamma$ should scale as $\sim g^2E^3$.
This can be confirmed by writing down the decay rate of the Tkachenko mode:
\begin{multline}\label{decay}
  \Gamma_\q = \frac1{2\epsilon_\q} \frac12 \!\int \!
  \frac{d^2\p}{(2\pi)^2 2\epsilon_\p 2\epsilon_{\q-\p}}\,
  |\mathcal M(\q\to \p, \q-\p)|^2\times\\
  \times (2\pi)\delta(\epsilon_\q-\epsilon_\p-\epsilon_{\q-\p}).
\end{multline}
Estimating the integral with $p\sim q$, $\mathcal M\sim g q^6$, we get $\Gamma_\q \sim g^2q^6\sim g^2E^3$. The presence of an anisotropic cubic vertex means that the decay rate depends on the direction of the momentum of the decaying particle.

At small but finite temperature $T$, the $U(1)$ condensate phase disappears, but the order parameter of translation symmetry breaking $\d_i\phi$ has a logarithmic correlation at long distances~\cite{Baym:2004} (see also Refs.~\cite{Lake:2022ico,Kapustin:2022fzp}).  Below the Berezinskii-Kosterlitz-Thouless phase transition where the lattice melts, the Tkachenko mode should still exist.  The $1\to 2$ decay rate~(\ref{decay}) is modified for modes with energy much less than $T$
by the factor $(1+f_\p+f_{\q-\p})$ where $f_\p$ and $f_{\q-\p}$ are the occupation numbers in the final state.
%
For $E\ll T$, 
this factor is of order $T/E$, hence the $1\to2$ rate is now $g^2TE^2$ for $E\ll T$.  However, the dominant contribution to the width is now a different process: the ``Landau damping'' process, i.e., the absorption of the soft Tkachenko quantum by a hard thermal Tkachenko photon in the medium:
\begin{multline}
  \Gamma_\q = \frac1{2\epsilon_\q} \!\int \!
  \frac{d^2\p}{(2\pi)^2 2\epsilon_\p 2\epsilon_{\q+\p}}\,
  |\mathcal M(\q, \p \to \q+\p)|^2\times \\
  \times 
  (f_\p - f_{\q+\p})
  (2\pi)\delta(\epsilon_\q{+}\epsilon_\p{-}\epsilon_{\q+\p}).
\end{multline}
The width of the Tkachenko mode due to this process is $g^2(TE)^{3/2}$, which means that the Tkachenko mode is still a well-defined resonance.

The estimate above assumes that the hard Tkachenko quanta participating in the scattering process has no width and is valid only when the energy of the Tkachenko mode under consideration is larger than the width of a typical thermal mode, which is, by dimensional analysis, $g^2T^3$.
Thus the estimate $\Gamma(E)\sim g^2(TE)^{3/2}$ is valid in the interval $g^2T^3\ll E\ll T$.  The regime $E\ll g^2T^3$ is the hydrodynamic regime, the analysis of which we defer to future work.

We note that our formulas for the width of the Tkachenko mode, both at zero and nonzero temperature, are in conflict with a previous result obtained from a microscopic calculation~\cite{Matveenko:2011}.  For bosons on the LLL, the authors of Ref.~\cite{Matveenko:2011} found that at zero temperature ratio of the width to the energy of the Tkachenko mode is a constant independent of the energy (which depends only on the filling factor), and at nonzero temperature the mode is overdamped.  The results are untypical for a NGB, and we cannot reconcile them with the symmetries of the system.  This discrepancy needs to be investigated further.

\emph{Conclusion.}---In this Letter, we have provided a new interpretation of the Tkachenko mode in a rotating superfluid: it is a noncommutative Nambu-Goldstone boson that arises from the breaking of $U(1)$ and translation symmetries.  Noncommutative field theory provides a convenient way to impose the invariance of the theory with respect to $U(1)$ and magnetic translations, and the resulting theory gives us a prediction for the decay width of the Tkachenko mode at low momentum.

\acknowledgments

The authors thank Clay C\'ordova, Austin Joyce, Umang Mehta, Nathan Seiberg, Lev Spodyneiko, and Wilhelm Zwerger for discussions and comments. S.M. is supported by Vetenskapsr\aa det (grant number 2021-03685) and Nordita. The work of D.X.N. is supported, in part, by Grant IBS-R024-D1. This work of D.T.S. and Y.-H.D. is supported, in part, by the U.S.\ DOE Grant No.\ DE-FG02-13ER41958, a Simons Investigator Grant (DTS) and by the Simons Collaboration on Ultra-Quantum Matter, which is a Grant from the Simons Foundation (No.\ 651440, DTS). 

\bibliography{LLLVL.bib}

\clearpage
\newpage
\begin{widetext}
	\begin{center}
		\textbf{\large  --- Supplementary Material ---\\ $~$ \\
			Noncommutative Field Theory of the Tkachenko Mode: Symmetries and Decay Rate}\\
		\medskip
		\text{Yi-Hsien Du, Sergej Moroz, Dung Xuan Nguyen, and Dam Thanh Son}
		
	\end{center}
	\setcounter{equation}{0}
	\setcounter{figure}{0}
	\setcounter{table}{0}
	\setcounter{page}{1}
	\makeatletter
	\renewcommand{\thesection}{S\arabic{section}}
	\renewcommand{\theequation}{S\arabic{equation}}
	\renewcommand{\thefigure}{S\arabic{figure}}
	\renewcommand{\bibnumfmt}[1]{[S#1]}

\section{Redundancies of spontaneous broken symmetries}
We believe that the most transparent understanding of spontaneous symmetry breaking and the Goldstone boson counting for a two-dimensional superfluid vortex crystal was obtained by Watanabe and Murayama in Ref. \cite{PhysRevLett.110.181601}. Here we summarize their explanation adopted to the lowest Landau level regime.

Emergence of a vortex lattice ground state in a rotating superfluid breaks spontaneously global particle number $U(1)$ symmetry, magnetic translation symmetry, and magnetic rotation symmetry.  Notwithstanding, the generators of all these symmetries are linearly related to each other.  In particular, the momentum density $T^{0i}$ is given by
\begin{equation}
	\label{eq:Ti}
	T^{0i}=m j^i -B \epsilon^{ij}x_j n
\end{equation}
where $j^i$ and $n$ are the boson current and particle densities, respectively, and $B$ is the effective magnetic field originating from the rotation \footnote{ One can understand the last term of \eqref{eq:Ti} as the contribution from the Lorentz force. If we take the time derivative of that equation and use the continuity equation $\dot{n}=-\partial_i j^i$, we have the time variation of the total momentum is given by
	\begin{equation}
		\int d^2 x \,\dot{T}^{0i}=\int d^2 x \,\frac{\partial(m j^i)}{\partial t}- \int d^2 x \, \epsilon^{ik} j^k B, 
	\end{equation}
	the last term is nothing but the total Lorentz force acting on the system.}. In the massless regime $m\to 0$, where only the lowest Landau level states survive, we can ignore the first term in Eq. \eqref{eq:Ti}, so the momentum density operator and the particle density operator are proportional to each other. Furthermore, we can define the angular momentum density as $\mathcal{J}= \epsilon^{ij} x^i T^{0j}$
which is also simply related to the boson density operator $\mathcal{J}= B \vec{x}^2 n$. As a result, the densities of all symmetries that are spontaneously broken are not independent, but are linearly related to each other. Therefore we only have a single Goldstone boson, which is the Tkachenko mode.

\section{Linearized theory of the Tkachenko mode}
\label{sec:SM-linear}

\subsection{Linearized effective Lagrangian}
Our departure point is the low-energy linearized effective theory of a two-dimensional superfluid vortex lattice introduced in Refs.~\cite{MorozVL:2018, PhysRevLett.122.235301}.  We consider a system of bosons with density $n_0$ placed in a constant magnetic field $B$.  This magnetic field may be effectively created by rotating the system with angular frequency $B/(2m)$, at the same time putting it in a harmonic trap with the trap frequency fine-tuned to cancel the centrifugal force. The lattice is parametrized by the displacement field $u^i$, $i=x,y$, while the superfluid is characterized by the dual $u(1)$ gauge field $a_\mu$.
The Tkachenko mode emerges as the result of the mixing between the of elastic waves on the vortex lattice and the superfluid fluctuations.
We start from the leading-order (LO) quadratic Lagrangian linearized around the vortex crystal ground state \footnote{The normalization of the dual gauge field $a_{\mu}$ differs here by a factor $2\pi$ from the normalization used in Refs.~\cite{MorozVL:2018, PhysRevLett.122.235301}. Moreover, to be consistent with the main text, the sign of the $U(1)$ source $\mathcal{A}$ also differs with Refs.~\cite{MorozVL:2018, PhysRevLett.122.235301}.} 

\begin{equation} \label{EFT_lin}
\mathcal{L}^{(2)}=-\frac{B n_{0}}{2} \epsilon_{i j} u^{i} \dot{u}^{j}+\frac{B}{2\pi} e_{i} u^{i}-\frac{\lambda}{2} \frac{\delta b^{2}}{(2\pi)^2} +\frac1{2\pi}\epsilon^{\mu\nu\rho} \mathcal{A}_\mu \partial_\nu a_\rho -\mathcal{E}_{\mathrm{el}}^{(2)}(\partial u).
\end{equation}
The formula for the particle number spacetime current in terms of the gauge field, $j^\mu=\delta S/\delta \mathcal{A}_\mu=\frac1{2\pi}\epsilon^{\mu\nu\rho} \partial_\nu a_{\rho}$, relates the particle number density $n$ with the magnetic field $b$ and the particle number current with the electric field 
$e_i=\partial_t a_i-\partial_i a_t$.
The term with one time derivative, proportional to $\epsilon_{ij}u^i \dot u^j$, encodes the Berry phase that a vortex acquires when moving in a superfluid.  This term gives rise to the ``Magnus force'' acting on the vortex.  The elastic energy density $\mathcal{E}_{\mathrm{el}}^{(2)}(\partial u)$ is a function of the linearized strain tensor $u_{i j}=\frac12\left(\partial_i u_j+\partial_j u_i\right)$. The superfluid internal energy is a function of the superfluid density $n=\frac1{2\pi}b$ and here is expanded around the ground state value $n_0$ to quadratic order in fluctuations $\delta b=b-2\pi n_0$. Finally, we included the coupling to an external $U(1)$ source $\mathcal{A}_\mu$ which is set to vanish in the ground state \cite{MorozVL:2018}.


The quadratic Lagrangian \eqref{EFT_lin} can be easily obtained from the Lagrangian \eqref{eq:L0} in the main text \footnote{Notice, that the elastic term was for simplicity suppressed in Eq. \eqref{eq:L0}, but can be easily expressed in terms of the fields $X^a(t,x^i)$, see e.g. Appendix B of Ref. \cite{MorozVL:2018}}. To this end following \cite{MorozVL:2018}, we substitute into Eq. \eqref{eq:L0} the definition of the vortex current \eqref{eq:jv}, and the definition of $X^a$ in terms of $u^a$, see Fig. \ref{FS1}. We then expand the action up to the quadratic order in the field fluctuations and arrive at the linearized action \eqref{EFT_lin} with the fluctuation of the vector potential source $\mathcal{A}_\mu=A_\mu- \bar{A}_\mu$ on top of the background $\bar{A}_\mu$ that produces the constant background magnetic field $B$.

\begin{figure}
\centering
\includegraphics[width = 0.5\columnwidth]{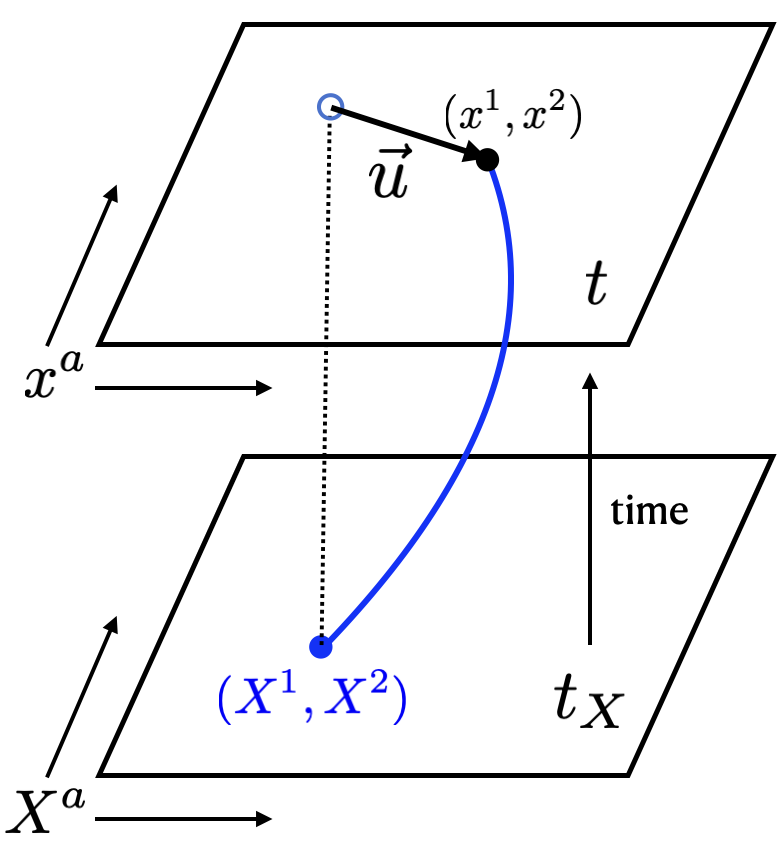}
\caption{The cartesian coordinates $X^1$ and $X^2$ label vortices at the initial time $t_{X}$. These coordinates are frozen into the vortex system and thus are fixed along each vortex (blue) worldline during time evolution. The displacement $u^a(t, x^i)=x^a-X^a(t, x^i)$. }
\label{FS1}
\end{figure}

In the absence of the $U(1)$ source, one finds after integrating out $a_i$ (see Appendix of Ref.~\cite{Jeevanesan:2022} for details)
\begin{equation} \label{Sint}
S_{\mathrm{eff}}\left[u^{i}, a_{t}\right]=\int d t d \mathbf{x}\left(\mathcal{L}_{\mathrm{el}}^{(2)}+\frac{B}{2\pi} a_{t} \partial_{i} u^{i}\right)+\frac{B^{2}}{2 \lambda} \int \frac{d t d \mathbf{k}}{(2 \pi)^{2}} \frac{\dot{u}_{-\mathbf{k}}^{i} \dot{u}_{\mathbf{k}}^{i}}{\mathbf{k}^{2}},
\end{equation}
where the first two terms in the Lagrangian \eqref{EFT_lin} constitute $\mathcal{L}_{\mathrm{el}}^{(2)}$.
Now one can integrate out the temporal component $a_t$ which gives rise to the Gauss law constraint $\partial_i u^i=0$. In the low-frequency and long-distance domain, the vortex crystal thus appears to be incompressible. This constraint can be resolved explicitly by introducing a dimensionless scalar field $\phi$ in terms of which $u^i=-\theta\epsilon^{ij}\partial_j \phi$, where following the main text we defined $\theta=-l^2=-1/B$. The transverse phonon $\phi$, known as the Tkachenko mode, is the only low-energy dynamical excitation mode of the vortex lattice.
In terms of the field $\phi$, the effective theory reduces to the local form
\begin{equation} 
\label{eq:Lphi}
\mathcal{L}_{\phi}=\frac{1}{2 \lambda}\dot \phi^{2}-\frac{C_{2}}{B_{0}^{2}}(\Delta \phi)^{2}.
\end{equation}
In contrast to phonons in ordinary crystals, the Tkachneko mode has a soft quadratic dispersion relation at low momenta.

\subsection{Coupling to $U(1)$ source and linear response}

After including the coupling to the $U(1)$ source via the mixed Chern-Simons term, we find that after integrating out $a_i$ one ends up with the action \eqref{Sint} with the replacement 
\begin{equation}
\dot{u}_{\mathbf{k}}^{i}\to \dot{u}_{\mathbf{k}}^{i}+\frac{\epsilon^{ij} \mathcal E^j_{\mathbf{k}}}{B}
\end{equation}
in the second integral. So the displacement velocity is measured with respect to the LLL drift that has the velocity $ -\epsilon^{ij} \mathcal E_j/ B$. As a result, the action is invariant under Galilean boosts. 
We thus have the effective action
\begin{equation} \label{SintS}
\begin{split}
	S_{\mathrm{eff}}\left[u^{i}, a_{t}\right]=&\int d t d \mathbf{x}\left(\mathcal{L}_{\mathrm{el}}^{(2)}+\frac{a_{t}}{2\pi} [B \partial_{i} u^{i}+\mathcal B]\right)+\\
	&\frac{B^{2}}{2 \lambda} \int \frac{d t d \mathbf{k}}{(2 \pi)^{2}} \left(\dot{u}_{-\mathbf{k}}^{i}+\frac{\epsilon^{ij} \mathcal{E}^j_{-\mathbf{k}}}{B_0} \right) \frac{1}{\mathbf{k}^{2}} \left(\dot{u}_{\mathbf{k}}^{i}+\frac{\epsilon^{ij} \mathcal{E}^j_{\mathbf{k}}}{B_0} \right),
\end{split}
\end{equation}
where $\mathcal{B}=\epsilon^{ij}\partial_i \mathcal{A}_j$ is a variation of the magnetic field on top of the constant background $B$.
In the presence of a such inhomogenity, the Gauss law is $\partial_i u^i=  -\mathcal B/B$, so the crystal becomes compressible. In momentum space (our conventions: $\partial_i\to i k_i$ and $\partial_t\to - i \omega$), this Gauss law can be resolved as
$u_{\mathbf{k}}^{i}= -\theta \epsilon^{i}_{\,\, j} (i k^j \phi_{\mathbf{k}}- \mathcal A^j_{\mathbf{k}})$.
The way the dimensionless scalar $\phi$ couples to the $U(1)$ source suggest that in addition to fixing the transverse displacement of the vortex crystal, $\phi$ also represents the regular part of the superfluid phase of the Bose-Einstein condensate. The latter interpretation was the key point for
Watanabe and Murayama, who developed the effective theory of the superfluid vortex crystal in Ref.~\cite{PhysRevLett.110.181601}.

Now we are ready to write the generalization of the effective theory \eqref{eq:Lphi} in the presence of the $U(1)$ source.
The simplest result is obtained, when one considers a special type of the source with vanishing $\mathcal B$ which thus does not violate the incompressibility condition. To this end, we will set $\mathcal A_i=0$. In that case, after resolving the Gauss law, we find the quadratic effective action for the $\phi$ fluctuation

\begin{equation}
\begin{split} \label{SA0}
	S_{\mathrm{eff}}\left[\phi\right]
	&=\frac{1}{2 \lambda} \int \frac{d \omega d \mathbf{k}}{(2 \pi)^{3}}\left(( -i \omega \phi_{-k}+\mathcal A^0_{-k}) (i \omega \phi_{k}+\mathcal A^0_k) -\frac{2 C_{2} \lambda}{B^{2}} \phi_{-k} \mathbf{k}^4 \phi_{k}  \right),
\end{split}
\end{equation}
where $k=(\omega, \mathbf{k})$.

We will extract now the density susceptibility $\chi_k$ that (up to a sign) is just the correlation function $\langle n_{-k} n_k\rangle$. To this end, we first compute the superfuid density
\beq \label{neq}
n_k=\frac{\delta S}{\delta \mathcal A^0_{-k}}=\frac{1}{\lambda}\left( i \omega \phi_{k}+ \mathcal A^0_k \right)
\eeq
substitute into it the solution of the equation of motion for $\phi_{-k}$
\begin{equation}
\phi_k=\frac{i \omega \mathcal A_k^0}{\omega^2-\frac{2 C_{2} \lambda}{B^2} \mathbf{k}^4}
\end{equation} 	
and get
\beq
n_k=-\frac{2 C_2 \mathbf{k}^4 \mathcal A_k^0}{B^2 (\omega^2-\frac{2 C_{2} \lambda}{B^2} \mathbf{k}^4)}.
\eeq
Finally, we differentiate the result with respect to $\mathcal A_k^0$ to get
\beq
\chi_k=-\frac{\partial n_k}{\mathcal A_k^0}=\frac{2 C_2 \mathbf{k}^4}{B^2 (\omega^2-\frac{2 C_{2} \varepsilon''}{B^2} \mathbf{k}^4)}.
\eeq
This agrees with the result of Ref.~\cite{MorozVL:2018}. 

More generally, the linear electromagnetic responses extracted from the effective action for the Tkachenko field $\phi$ are expected to agree with the massless LLL limit of the results derived in Ref. \cite{MorozVL:2018}. This amounts to discarding the contribution originating from the Kohn's mode \footnote{Notice that in Ref.~\cite{MorozVL:2018}, the contribution of the Kohn's mode is higher order in the derivative expansion since the power-counting scheme with $\omega \sim \mathbf{k}^2$ was used there. That counting originates from the dispersion of the low-energy Tkachenko mode.}.


\subsection{Hamiltonian formulation} \label{sec:Haml}
Starting from the Lagrangian \eqref{eq:Lphi}, the canonical momentum conjugate to $\phi$ is
\beq
\label{eq:Haml}
\pi_\phi=\frac{\partial \mathcal{L}_\phi}{\partial \dot \phi}=\frac{1}{\lambda}\dot \phi
\eeq
which according to Eq. \eqref{neq} is (minus) the superfluid density $\pi_\phi=-n$. So $\phi$ is indeed the superfluid phase field.  The Hamiltonian density can now be computed to be
\beq
\mathcal H= \pi_\phi \dot \phi- \mathcal{L}_{\phi}=\frac 1 2 \lambda \pi_{\phi}^2+ \frac{C_2}{B^2} (\Delta \phi)^2.
\eeq
Using now the canonical Poisson bracket $[\phi(\mathbf{x}), \pi_{\phi}(\mathbf{y}) ]=\delta(\mathbf{x}-\mathbf{y})$, we end up with the Hamiltonian equations of motion
\beq
\begin{split}
\partial_{t} \phi&=[\phi, H]=\lambda \pi_{\phi}=-\lambda n, \\
\partial_{t} \pi_{\phi}&=-\partial_{t} n=[\pi_{\phi}, H]=-2 \frac{C_{2}}{B^2} \Delta^{2} \phi.
\end{split}
\eeq
We observe that the time-evolution of the Tkachenko field $\phi$ is fixed by the superfluid density, while the time-evolution of the latter is fixed by the fourth-spacial derivative of $\phi$.

\section{Resolution of the nonlinear constraint}
\label{sec:SM-nonlinear}
\subsection{Constant magnetic field}
As argued in the main text, in a constant effective magnetic field $B=2m \Omega$, 
the scalar fields $X^a$ must satisfy the non-linear constraint
\begin{equation}\label{eq:constraint1}
\frac{1}{2} \epsilon^{ij}\epsilon^{ab}\d_i X^a \d_j X^b=1.
\end{equation}
To resolve the constraint, introduce an {\it auxiliary Poisson bracket} 
\begin{equation}
\{f,g\}=\epsilon^{ij} \d_i f \d_j g
\end{equation}
such that
\begin{equation}
\label{eq:Poisson}
\{x^a,x^b\}=\epsilon^{ab}.
\end{equation}
As the consequence of the constraint \eqref{eq:constraint1}, the fields $X^a$ also satisfy
\begin{equation}
\label{eq:Poisson1}
\{X^a,X^b\}=\epsilon^{ij} \frac{\d X^a}{\d x^i}\frac{\d X^b}{\d x^j}=\epsilon^{ab}.
\end{equation}
Thus the transformation from $x^i$ to $X^a$ belongs to the group of canonical transformation and can be generated by a scalar function $\phi$ \footnote{The argument we use here is nothing but saying that $X^a$ and $x^i$ are related by a area-preserving diffeomorphism generated by a function $\phi(\mathbf{x})$. Notice that the auxiliary Poisson bracket in Eq.~\eqref{eq:Poisson} does not originate from the effective field theory Lagrangian. } 
\begin{equation}
\label{eq:X}
\begin{split}
	X^a&=x^a- \theta  \{\phi,x^a \}+\frac{\theta^2}{2!} \{ \phi,\{\phi,x^a\}\}+ \dots \\
	&=x^a+ \theta  \epsilon^{ab}\d_b \phi-\frac {\theta^2} {2} \epsilon^{ab} \epsilon^{cd} \partial_c \phi \partial_b \partial_d \phi + \dots
\end{split}	
\end{equation}
Notice that this expression of $X^a$ is identical with Eq. \eqref{eq:transX} in the main text. As a superfluid order phase, the scalar generator $\phi$ is  even under 2d parity $x\leftrightarrow y$ and odd under time reversal $t\to -t$.

Here we comment on two natural ways how to organize the derivative expansion in terms of $\phi$: If we scale $\phi\sim O(\epsilon^{-1})$ and $\partial_i \sim O(\epsilon)$, higher order non-linearities in the expansion \eqref{eq:X} are systematically suppressed. On the other hand, in order to include all non-linearities on equal footing, we can scale $\phi\sim O(\epsilon^{-2})$ and $\partial_i \sim O(\epsilon)$. In this way, all non-linear terms in Eq. \eqref{eq:X} are of the same order.

\subsection{Inhomogeneous magnetic field} \label{sec:const_inhomB}
It is possible to generalize the above construction to the case, where the effective magnetic field is not constant. The constraint to be resolved is now
\begin{equation}\label{eq:constraintB}
\frac{1}{2} \epsilon^{ij}\epsilon^{ab}\frac{\d X^a}{\d x^i} \frac{\d X^b}{\d x^j}=\frac{B+\mathcal{B}}{B}=1-\theta\mathcal B,
\end{equation}
where $\mathcal B$ is a magnetic perturbation on top of a constant background $B$. We resolve the constraint  \eqref{eq:constraintB} by the following ansatz 
\begin{equation}
X^a=x^a+y^a(x)
\end{equation}
$y^a$ is a perturbation in the same order as the perturbed background fields. We write the shift $y^a$ in perturbative orders 
\begin{equation}
y^a=y^a_1 + y^a_2 + \cdots
\end{equation}
From the previous derivation, we can easily guess the first order 
\begin{equation}
\label{eq:y1}
y^a_1=-\theta[\{ \phi(x),x^a\}+ \epsilon^{ab}\mathcal{A}_b]=\theta\epsilon^{ab}\underbrace{\left(\d_b \phi-\mathcal{A}_b\right)}_{D_b\phi},
\end{equation}
where the field $\mathcal{A}_b$ satisfies 
\begin{equation}
\epsilon^{ab} \d_a \mathcal{A}_b=\mathcal B.
\end{equation}
In the perturbative approach, we consider that $\phi$ and $\mathcal B$ are of the same order. 
Given that $\phi$ is the superfluid phase, the perturbation $y^a_i$ is invariant under a $U(1)$ gauge transformation 
\begin{equation}
\label{eq:gauge0}
\phi \to \phi + \beta, \quad \mathcal{A}_i \to \mathcal{A}_i+ \d_i \beta. 
\end{equation}

One can check that the constraint \eqref{eq:constraintB} can be satisfied up to all orders in perturbation if we choose the recurrence relation for $y^a_n$ as follows 
\begin{equation}
\label{eq:recurr}
y^a_n=-\sum_{0<m<n}\frac{1}{2} \epsilon^{ab} \epsilon^{cd} y^c_m \d_b y^d_{n-m}.
\end{equation}
Explicitly, the second order term $y^a_2$ is 
\begin{equation}
y^a_2=-\frac{1}{2} \epsilon^{ab} \epsilon^{cd} y^c_1 \d_b y^d_{1}
=-\frac{\theta^2}{2} \epsilon^{ab} \epsilon^{cd} D_c \phi \d_b D_d \phi.
\end{equation}
So given $y^a_1$ from Eq. \eqref{eq:y1}, we can, in principle, obtain $y^a_n$ for all orders $n$ using repeatedly the recurrence relation \eqref{eq:recurr}.

\section{Nonlinear theory of the Tkachenko mode from the EFT of Ref.~\cite{MorozVL:2018}}
\label{sec:SM-derivation}

We begin with the non-linear effective theory of the vortex lattice introduced in \cite{MorozVL:2018}
\begin{equation} \label{EFTLO}
\cL=-\frac{B}{4\pi}\epsilon^{\mu\nu\lambda}\epsilon^{ab}a_\mu \d_\nu X^a \d_\lambda X^b-\varepsilon(b)-\varepsilon_{el}(U_{ab})+ \frac1{2\pi}\epsilon^{\mu\nu\lambda}A_\mu \d_\nu a_\lambda.
\end{equation}
In this formulation, positions of vortices are encoded in two scalar fields $X^1(t, \mathbf{x})$ and $X^2(t, \mathbf{x})$ that are the Lagrange coordinates frozen into the vortex lattice. The first term in the Lagrangian \eqref{EFTLO} encodes the coupling of the vortex current to the dual $u(1)$ gauge field. Given that in the following we want to integrate out superfluid density fluctuations, we will expand the energy density $\varepsilon(b)$ around its minimum $b=b_0$ and keep track only of the quadratic term  
\begin{equation}
\label{eq:eb}
\varepsilon(b)= \varepsilon_0+\frac{\lambda}{2}\frac{\delta b^2}{(2\pi)^2}+\dots,
\end{equation} 
with $\delta b = b-b_0$. The elastic energy density $\varepsilon_{el}$ depends on $X^a$ fields via the combination 
$	U^{ab}=\delta^{ij}\d_i X^a \d_j X^b$ \cite{Leutwyler1996, soper2008}.
Finally, the superfluid current $j^\mu=\delta S / \delta A_\mu=\frac{1}{2\pi}\epsilon^{\mu\nu\lambda}\partial_{\nu}a_\lambda$ is coupled minimally to the external $U(1)$ source $A_\mu$ that includes the background magnetic field $B$. 

Following \cite{MorozVL:2018}, one can introduce a derivative expansion with the power-counting scheme
\begin{equation} \label{PC}
a_i, X^a, A_i \sim O\left(\epsilon^{-1}\right), \quad a_t, A_t \sim O\left(\epsilon^0\right), \quad \partial_i \sim O\left(\epsilon^1\right), \quad \partial_t \sim O\left(\epsilon^2\right),
\end{equation}
where $\epsilon$ is a small parameter. The difference in scaling of time and spatial derivatives has its root in the quadratic dispersion of the collective Tkachenko mode. Within this derivative expansion, all terms in the Lagrangian \eqref{EFTLO} are of order $\epsilon^0$ and will be called leading-order (LO) in the following. Higher derivative next-to-leading (NLO) corrections to these terms have been considered in the literature \cite{MorozVL:2018, PhysRevLett.122.235301} and we will discuss them briefly in Sec. \ref{Sec:NLO}.


\subsection{Effective action} \label{sec:EFA}
We use the resolution of the non-linear constraint discussed in Sec. \ref{sec:SM-nonlinear} and formulate the LO non-linear effective theory in terms of the scalar field $\phi$. Here we will derive explicitly only leading order non-linearities that involve $\dot \phi$, a more general discussion (based on symmetries) of allowed non-linear terms can be found in the main text. We will turn off $\mathcal A_i$ perturbation on top of the constant magnetic background, but will keep arbitrary $A_0$.
The vortex current which couples minimally to the dual gauge field $a_i$ is 
\begin{equation}
\label{eq:Ji}
j^i_v=\frac{B}{4\pi}\epsilon^{ij}\epsilon^{ab}\left( - \d_t X^a \d_j X^b+ \d_j X^a \d_t X^b\right).
\end{equation}
We then substitute the expression of $X^a$ \eqref{eq:X} into it and obtain the vortex current up to second order in $\phi$
\begin{equation}
j^i_v=\frac{1}{2\pi}\epsilon^{ij} \d_j (\dot{\phi}+\frac{\theta}{2} \epsilon^{kl}\d_k \dot{\phi} \d_l \phi)+O(\phi^3).
\end{equation}
Now we will rewrite the theory \eqref{EFTLO} in terms of the field $\phi$
\begin{equation}
\cL=-\frac{1}{2\pi} a_i \epsilon^{ij} \d_j (\dot{\phi}+\frac{\theta}{2} \epsilon^{kl}\d_k \dot{\phi} \d_l \phi)-\frac{\lambda}{2}\frac{(\epsilon^{ij}\d_i \delta a_j)(\epsilon^{kl}\d_k \delta a_l)}{(2\pi)^2}-\varepsilon_{el}(U^{ab})+\frac{1}{2\pi}A_0 \epsilon^{ij}\d_i a_j,
\end{equation} 
with the fluctuation of the emergent gauge field defined by $\epsilon^{ij}\d_i\delta a_j=\delta b$.  The equation of motion of $\delta a_i$ is the constraint 
\begin{equation}
-\epsilon^{ij} \d_j (\dot{\phi}+\frac{\theta}{2} \epsilon^{kl}\d_k \dot{\phi} \d_l \phi)-\frac{\lambda}{2\pi} \epsilon^{ij} \d_j \delta b+\epsilon^{ij} \d_j A_0=0
\end{equation}
with the solution
\begin{equation}
\delta b=-\frac{2\pi}{\lambda}\left(\dot{\phi}+\frac{\theta}{2} \epsilon^{kl}\d_k \dot{\phi} \d_l \phi-A_0 \right).
\end{equation}
Substituting this into the Lagrangian gives us 
\begin{equation} \label{EFTphi}
\cL=\frac{1}{2 \lambda} \left(\dot{\phi}+\frac{\theta}{2} \epsilon^{kl}\d_k \dot{\phi} \d_l \phi-A_0\right)^2-\frac{b_0}{2\pi}\left(\dot{\phi}+\frac{\theta}{2} \epsilon^{kl}\d_k \dot{\phi} \d_l \phi-A_0\right)-\varepsilon_{el}(U^{ab}).
\end{equation}
The form of the coupling between the scalar $\phi$ and $A_0$ is fixed by Galilean symmetry \footnote{While the coupling to $A_0$ source is local, one can check that in this formulation the coupling to the $\mathcal A_i$ source is non-local.}.
This Lagrangian is the leading non-linear generalization of the effective theory \eqref{eq:Lphi}, i.e., it captures reliably the  cubic non-linear dynamical term $\frac{\theta}{2\lambda} \dot \phi \epsilon^{kl}\partial_k \dot \phi \partial_l \phi$ and modifies the coupling of the field $\phi$ to the potential $A_0$. While the former is a total derivative and does not change the equations of motion, the latter modifies the expression of the $U(1)$ particle number density in terms of $\phi$ and gives rise to the celebrated GMP algebra, see Sec. \ref{sec:GMP}. Furthermore, one can see that the dynamical terms in the Lagrangian \eqref{EFTphi} that are proportional to $b_0$ can be rewritten as total derivatives and thus do not affect the equations of motion. 


The boson particle number density is 
\begin{equation}
\label{eq:density}
n=\frac{\delta S} { \delta A_0}=\frac{b_0}{2\pi}-\frac{1}{\lambda}\left[\dot{\phi}+\frac{\theta}{2} \epsilon^{kl}\d_k \dot{\phi} \d_l \phi-A_0\right]
\end{equation}
with the background bosonic charge $n_0=b_0/(2\pi)$ and the fluctuation $\delta n$ fixed by the field $\phi$. Notably, the first two dynamical terms of the action \eqref{EFTphi} originate from the short-range interaction between the elementary bosons, namely the first term is just 
$
\frac{\lambda}{2} \delta n^2
$,
while the second term of \eqref{EFTphi} is $\lambda n_0 \delta n$.


Finally, we work out the canonical structure of the theory \eqref{EFTphi}.
The canonical conjugate momentum of $\phi$ is
\begin{equation}
\pi_\phi=\frac{\delta \cL}{\delta \dot{\phi}}= \frac{1}{\lambda} \dot{\phi}+\cO(\phi^2).
\end{equation}	
Given the canonical commutation relation 
\begin{equation}
[\phi(\x),\pi_\phi(\y)]=i \delta(\x-\y),
\end{equation}	 
at second order we end up with the leading order canonical commutation relation \footnote{The quadratic in $\phi$ contribution to the canonical conjugate momentum $\pi_\phi$ does not contain temporal derivatives and thus does not affect the canonical commutation relation \eqref{eq:canon} up to second order in $\phi$. }
\begin{equation}
\label{eq:canon}
[\phi(\x),\dot{\phi}(\y)]=i \lambda \delta(\x-\y)
\end{equation}	
which is identical to the linearized theory, see Eq. \eqref{eq:Haml}.
\subsection{GMP algebra} \label{sec:GMP}
The particle number density, in the absence of the background $A_0$, extracted from the non-linear effective theory \eqref{EFTphi} is given by 
\begin{equation}
\label{eq:rho}
n=\frac{b_0}{2\pi}-\frac{1}{\lambda}(\dot{\phi}+\frac{\theta}{2} \epsilon^{kl}\d_k \dot{\phi} \d_l \phi)+O(\phi^3).
\end{equation}
In the derivation of the GMP algebra, we can ignore the background part of the charge density $n_0=b_0/(2\pi)$ since it only contributes to the zero momentum density, which is absent in the GMP algebra. Here we will compute explicitly the commutation relation of particle number density operator 
\begin{equation}
[n(\x),n(\y)]=\frac{1}{\lambda^2}\left[\dot{\phi}(\x)+\frac{\theta}{2} \epsilon^{kl}\d_k \dot{\phi}(\x) \d_l \phi(\x),\dot{\phi}(\y)+\frac{\theta}{2} \epsilon^{mn}\d_m \dot{\phi}(\y) \d_n \phi(\y)\right].
\end{equation}
To this end, we use the commutation relation \eqref{eq:canon} and obtain 
\begin{multline}
\label{eq:GMPx}
[n(\x),n(\y)]=\frac{1}{\lambda}\left[ \frac{\theta}{2} \epsilon^{kl} \frac{\d}{\d x^k} \dot{\phi}(\x)\frac{\d}{\d x^l} i \delta(\x-\y)-\frac{\theta}{2} \epsilon^{mn} \frac{\d}{\d y^m} \dot{\phi}(\y)\frac{\d}{\d y^n} i \delta(\y-\x)\right. \\
\left. +\frac{\theta^2}{4} \epsilon^{kl}\epsilon^{mn}\frac{\d}{\d x^k} \dot{\phi}(\x)\frac{\d}{\d y^n} \phi(\y) \frac{\d}{\d x^l}\frac{\d}{\d y^m}i \delta(\x-\y)-\frac{\theta^2}{4} \epsilon^{kl}\epsilon^{mn}\frac{\d}{\d y^m} \dot{\phi}(\y)\frac{\d}{\d x^l} \phi(\x) \frac{\d}{\d x^k}\frac{\d}{\d y^n}i \delta(\y-\x)\right].
\end{multline}
Now we do Fourier transformation in both $\x$ and $\y$ by taking the integral 
$	\int d^2 \x d^2\y \,e^{i \mathbf{k} \x}e^{i \mathbf{q} \y} [\cdots]
$.
The first term of \eqref{eq:GMPx} gives us 
$	\frac{i\theta}{2\lambda} \k \times \q\dot{\phi}(\k+\q)
$.
The second term gives rise to an identical result. 
The summation of the third and the fourth terms produces 
\begin{equation}
-\frac{i \theta^2}{2\lambda} (\k\times \q)\int d^2 \x e^{i(\mathbf{k+q})\x} \epsilon^{ij}\frac{\d}{\d x^i} \dot{\phi}(\x) \frac{\d}{\d x^j} \phi(\x).
\end{equation} 
We combine the results of the Fourier transformation of \eqref{eq:GMPx} and obtain
\begin{equation} \label{eq:GMP}
[n(\k),n(\q)]=-i \theta (\k\times \q) n(\k+\q)=i \ell^2(\k\times \q) n(\k+\q)
\end{equation}
with the definition $\ell=1/\sqrt{B}$. We thus end up with the long wavelength version of the celebrated GMP algebra \cite{GMP} that indicate that our starting point \eqref{EFTLO} is a theory operating purely in the lowest Landau level. The long wavelength limit of GMP algebra was also obtained in the composite fermion theories \cite{Nguyen:DCF, Nguyen:DCFeven} and the bi-metric theory of fractional quantum Hall \cite{Gromov:Bimetric,Nguyen:Bimetric}.

\subsection{LLL volume-preserving diffeomorphisms}
Given that we work in the LLL regime, we consider here a combination of an infinitesimal two-dimensional volume-preserving diffeomorphism generated by
\begin{equation} \label{tr1}
x^i \to x^i+ \xi^i, \quad \xi^i=-\theta \epsilon^{ij}\d_j \alpha
\end{equation}
and a $U(1)$ gauge transformation 
\begin{equation} \label{tr2}
A_\mu \to A_\mu + \d_\mu \vartheta, \quad \vartheta= \alpha-\theta \epsilon^{ij} A_i \d_j \alpha.
\end{equation}
Under these transformations, in the constant magnetic field and in the LLL approximation, the background $A_0$ transforms as 
\begin{equation}
\delta_\alpha A_0 = \dot{\alpha}+ \theta \epsilon^{ij} \d_i A_0 \d_j \alpha=\dot{\alpha}+ \theta\{A_0, \alpha\},
\end{equation}
while $A_i$ is unchanged \cite{Du:2022}. The corresponding transformation of $\phi$ is given by
\begin{equation}
\delta_\alpha \phi =\alpha-\frac{\theta}{2} \{\phi, \alpha\}+ \cdots,
\end{equation}
where the non-linear terms originate from the non-commutativity of the area-preserving diffeomorphisms\footnote{Note that here we consider $\phi$ as the dynamical field generating an area-preserving diffeomorphism that captures the dynamical Tkachenko mode. On the other hand, $\alpha$ is the generator of a non-dynamical infinitesimal area-preserving diffeomorphism that parametrizes the local symmetry of the lowest Landau level system \cite{Du:2022}.}. It is straightforward to check that these transformations realize the canonical $w_{\infty}$ algebra on the Tkachenko field $\phi$, namely  $[\delta_{\alpha}, \delta_{\beta}]\phi=\delta_{\{\alpha, \beta\}_\theta}\phi$.

One can check that up and including the leading order non-linearity, the action built from the Lagrangian \eqref{EFTphi} is invariant under the combination of \eqref{tr1} and \eqref{tr2}. Indeed, the variation of the first and second terms of \eqref{EFTphi} is a total derivative \footnote{Indeed, the first term in the Lagrangian \eqref{EFTphi} is just the contact density-density interaction term and the charge density transforms under volume-preserving diffeomorphisms as $\delta n=1/B_0\epsilon^{ij} \d_i \alpha \d_j n$, which is a total derivative. Similarly, the variation of the second term is also a total derivative.}, while the elasticity energy density is invariant on its own. Notably, this invariance automatically insures the emergence of the LLL GMP algebra \cite{Du:2022} that we derived explicitly above. Moreover, it implies that in the LLL limit the charge current density can be expressed as a derivative of the stress tensor and the charge conservation law has a higher-rank form that arises naturally in higher-rank tensor gauge theories coupled to fractons \cite{Pretko:2016kxt, Pretko:2018a}.
\subsection{Beyond leading-order theory \eqref{EFTLO}} \label{Sec:NLO}
Following the power-counting scheme \eqref{PC}, one can systematically add sub-leading symmetry-allowed terms to the LO effective theory Lagrangian \eqref{EFTLO}. In fact, some next-to-leading (NLO) terms have already been investigated before.

In particular, already in Ref. \cite{MorozVL:2018}, the non-linear NLO term $m e_i^2/(4\pi b)$ (whose form is fixed by Galilean invariance) has been incorporated. Here $e_i=\partial_t a_i-\partial_i a_t$ is the dual electric field that encodes the superfluid current, and $m$ is the mass of the elementary boson. This term allows us to go beyond the LLL approximation and incorporate some effects of higher Landau levels into the low-energy description of the superfluid vortex lattice. In particular, the inclusion of this term gives rise to the finite-frequency Kohn mode in the EFT excitation spectrum, which correspondingly modifies the $U(1)$ linear response \cite{MorozVL:2018}.

Another NLO term that breaks time-reversal symmetry and has the form $e^i\partial_i b/(4\pi b)$ has been discussed in Ref. \cite{PhysRevLett.122.235301}. Given that this term does not depend on the mass $m$ of the boson, it survives in the LLL limit and incorporates higher-order corrections to the LLL coarse-grained description developed above. In particular, we expect that this term is responsible for leading-order non-linear corrections to the low-momentum GMP algebra \eqref{eq:GMP}. 

We expect that the NLO terms discussed above will generate corrections to the decay rate $\Gamma$ of the Tkachenko mode. Those however will disappear faster in the limit $E\to 0$ than the leading-order result $\Gamma\sim E^3$ that we discovered in this paper. 

Notice that either of the NLO terms mentioned here modify the Gauss law constraint \eqref{eq:constraint1} and make the vortex crystal compressible. As a result, the resolution of the constraint by a canonical transformation \eqref{eq:X} is not applicable anymore. Nevertheless, it should be possible to resolve the modified constraint by generalizing the method used in Sec. \ref{sec:const_inhomB}.

\end{widetext}

\end{document}